\documentclass[twocolumn,aps,floatfix,superscriptaddress]{revtex4}

\usepackage{graphicx,xcolor}
\usepackage{bm}

\usepackage{epsf}
\usepackage{amssymb}
\usepackage{wrapfig}

\usepackage{amsmath,amsfonts}

\newcommand{\bea}{\begin{eqnarray}}
\newcommand{\eea}{\end{eqnarray}}
\newcommand{\beq}{\begin{equation}}
\newcommand{\eeq}{\end{equation}}
\newcommand{\be}{\begin{equation}}
\newcommand{\ee}{\end{equation}}

\renewcommand{\>}{\rangle}
\newcommand{\<}{\langle}
\def\12{\frac{1}{2}}

\renewcommand{\phi}{\varphi}
\newcommand{\eq}[1]{(\ref{#1})}

\usepackage{hyperref}

\begin{document}
\title[Fermi distribution of coherent states \ldots]{Fermi distribution of semicalssical non-eqilibrium  Fermi states}

\author{E. Bettelheim}
\affiliation{Racah Institute of Physics, The Hebrew University of
Jerusalem, Safra Campus, Givat Ram, Jerusalem, Israel 91904.}


\author{Paul B. Wiegmann}
\affiliation{James Franck Institute
of the University of Chicago,
5640 S.Ellis Avenue, Chicago, IL 60637}

\begin{abstract}
When a classical device suddenly perturbs a degenerate Fermi gas a semiclassical non-equilibrium  Fermi state arises. Semiclassical Fermi states are characterized by a Fermi energy or Fermi momentum that slowly depends on space or/and time.     We show that the  Fermi distribution of a semiclassical Fermi  state has a universal nature. It  is described by Airy functions regardless of the details of the perturbation. In this letter we also  give a general discussion of  coherent Fermi states.
\end{abstract}
\date{\today}

\maketitle
\paragraph*{1. Introduction}
Among various excitations of  a degenerate Fermi gas, coherent Fermi states play a special role.
A typical (not coherent) excitation of a Fermi gas consists of a finite number of holes below the Fermi level and a finite number of particles above it. Instead, a coherent Fermi state  involves an infinite superposition of particles and holes arranged in such a manner  that one can still think in terms of a Fermi sea (with no holes in it), the Fermi level of which depends on time and space.

Coherent states  appear in numerous recent proposals about generating coherent quantum states in nanoelectronic devices and fermionic cooled atomic systems.
These states can be used to transmit quantum information, test properties of electronic systems and to generate many-particles entangled states.

Coherent Fermi states  can be obtained  by different means. One is a sudden perturbation of the Fermi gas. For example, a smooth potential well, the spatial extent of which much larger than the Fermi length, is applied to a Fermi gas.  Fermions are trapped in the well. Then the well is suddenly removed. An excited state of the Fermi gas obtained in this way is a coherent Fermi state. This kind of  perturbation is typical for various  manipulations with cooled fermionic atomic gases.

A realistic way to generate coherent Fermi  states in electronic systems is by  applying  a time dependent  voltage through a point contact, typifying many manipulations with nanoelectronic devices \cite{pointcontact,Keeling}.

Although the realization  of  coherent states in electronic systems experimentally  is more challenging  than in atomic systems, we will routinely talk about electrons.

From a theoretical standpoint coherent Fermi states are an  important concept revealing fundamental properties of Fermi statistics.
Coherent states appeared in other  disciplines not directly  related to electronic physics. Random matrix theory (RMT) \cite{Forrester-book}, non-linear waves  \cite{Miwa}, crystal growth \cite{Spohn,Okounkov}, various determinental  stochastic processes \cite{Johansson}, asymmetric diffusion processes \cite{TASEP},  to name but a few.

Unless a special effort is made (see e.g., \cite{pointcontact}) coherent Fermi states  involve many electrons and are such that  space-time gradients of the electronic density are much smaller than the Fermi scale. These state arise as a result of  perturbing tye Fermi gas by a classical device. We call them semiclassical Fermi states. They are the main object of this paper. We will show that semiclassical Fermi states show great degree of universality as well as their single electron counterpart studied in \cite{pointcontact,Keeling}.

A general  coherent Fermi  state is a unitary transformation of the ground state $|0\>$ of a Fermi gas:
$
|U\>=U|0\>,\quad U=e^{i\int \Xi (x)\rho(x)dx -i\int \Pi(x)v(x)dx},
$
where 
 $\rho$  and $v$ are operators of  electronic density and velocity $[\rho(x),\,v(y)]=-\frac \hbar m\nabla\delta(x-y)$ and  $\Xi(x)$ and $\Pi(x)$  are two real functions characterizing the state  \cite{Miwa,Perelomov}).

For the purpose of this paper it is sufficient to assume that the motion of electrons is one dimensional and chiral (electrons  move to the right), although most of the results we discuss are not limited to one-dimensional electronic gases.  To this end, edge states in the Integer Quantum Hall effect  may serve as a prototype. In  a chiral sector the  operators of density and velocity are identical $v|{\rm right}\>=\frac \hbar m\rho|{\rm right}\>$. The contribution of the right sector is easy to take into acoount.
 In this case the Fermi coherent state
 \begin{align}
 \label{P}|U\>=e^{i\int \Phi(x)\rho(x)dx}|0>
 \end{align}
 is characterized by a single  function $\Phi$.

The function  $\Phi$  can be undersstood as the action of an  instantaneous perturbation by a  potential $ \hbar v_F \nabla\Phi(x)$, or, if we ignore the electronic dispersion the function  $\Phi$ is the action of  a time-dependent gate voltage $eV(t)=-\hbar \frac{d}{dt}\Phi(x_0-v_Ft)$ applied through a point contact (located at $x_0$) \footnote{This is true before any gradient catastrophe had taken place, where  the momentum dependence of the velocity of the electrons can be ignored. The evolution of coherent states in dispersive Fermi gas was studied in \cite{ABWc}}.

\begin{figure}
\begin{center}
\includegraphics[width=8cm]{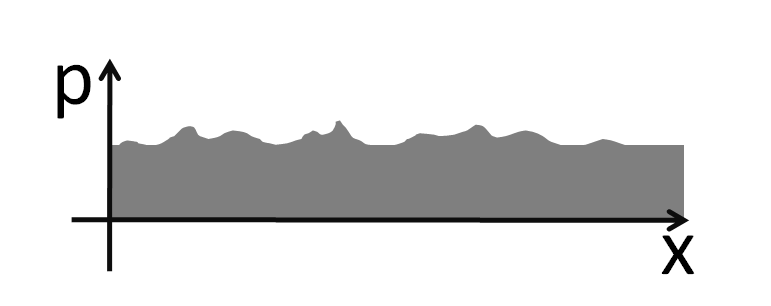}
\caption{ Modulated Fermi Edge. The Wigner function and Fermi number are almost 1 below the Edge (shaded area) and vanish above the Edge, having universal character in the 'surf'.  \label{Wigner}}
\end{center}
\end{figure}

Fermi coherent states feature inhomogeneous electric density $\rho(x)$ and current $I(x)$. Their expectation values  give a meaning to the functions $\Xi$ and $\Pi$. In the chiral state, where the electric current and the electronic density are proportional,  their expectation values are gradients of $\Phi$
 \begin{align}\label{ro}
\hbar\<U|\rho(x)|U\>=\frac{\hbar}{ e v_F} \<U|I(x)|U\>=p_F+\hbar\nabla\Phi.
 \end{align}
Consequently  $\hbar \nabla\Phi(x)$ plays the role of the space-time modulation of the Fermi point, such that all states with momentum (or energy) less than  $P_F(x)=\hbar \nabla\Phi(x)$ (or $E_F(x)=v_F \hbar \nabla\Phi$) are occupied (no holes in the Fermi sea) as shown in Fig.~\ref{Wigner}. We shall refer to the function $P_F(x)$ as the Fermi surface, with some abuse of nomenclature. We shall also term  the region in phase space around $P_F(x)$ as the 'Fermi surf'.

The question we address in this letter is: what is the Fermi distribution in the 'Fermi surf' of a modulated Fermi point, namely around and between extrema of $P_F(x)$?

We will be especially interested in semiclassical Fermi states. These states involve many excited electrons. Its typical range  is larger than the  momentum spacing  $\hbar \nabla  \Phi \gg \Delta$ (but still is smaller than the Fermi momentum).
\begin{figure}
\begin{center}
\includegraphics[width=8cm]{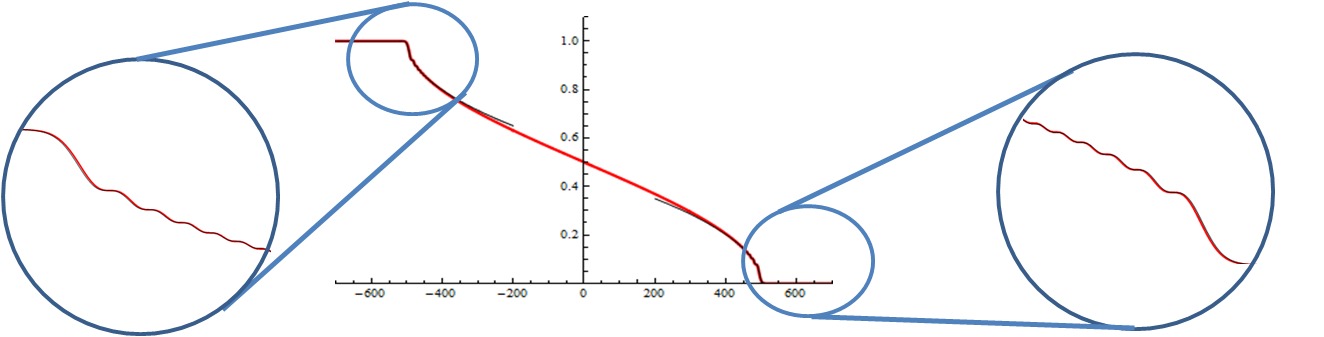}
\caption{Universal asymptotes of the Fermi number. The graph  (blue curve) shows the Fermi number  in units of $\hbar/\ell$ for the  example $\<\rho (x)\>=\rho_0+(n/\ell)\cos(x/\ell)$ computed from Eq.(\ref{BesselFermiNumber}). Black curves  are the asymptotic forms obtained from the universal Fermi number formula (\ref{W}).  The universal asymptotes are magnified.
 \label{TheFigure}}. \end{center}
\end{figure}

We  show  that, quite interestingly,  the  semiclassical Fermi surf features a universal Fermi distribution. There  the Wigner function (\ref{Wignerdefinition}) is described by the function ${\rm Ai}_1(s)=\int_s^\infty {\rm Ai}(s') ds'$:
\begin{align}\label{A}
n_F(x,p+P_F(x))\approx {\rm Ai}_1\left ( 2^{2/3} \kappa p\right),
\end{align}
where the scale  $\kappa=|\hbar^2P_F''(x_*)/2|^{-1/3}$, and the offset $P_F(x)$ are  the only information about the state that enters the formula. This formula holds close to any point where the Fermi surf is concave $P''_F(x)<0$.   The Fermi occupation number (\ref{FD})  also displays universal behavior. Near a maxima, but well above  the minima of the surf the Fermi number reads
\begin{align}\label{W}
n_F(P_F(x_*)+ p)\approx \kappa{\Delta}\left[ \left[{\rm Ai}'(\kappa p)\right]^2- (\kappa p) {\rm Ai}^2(\kappa p) \right].
\end{align}
If the Fermi surf is convex rather than concave, then particle hole symmetry $n_F(p,x) \to 1-n_F(-p,x)$ provides the result for the Fermi number and Wigner function.
Figures (\ref{Wigner},\ref{TheFigure}) illustrate the universal regimes. Here $\Delta$ is the momentum spacing ($2\pi \hbar/\Delta$ is the system volume).


The goal of the letter is twofold: to emphasize these simple, albeit universal,  distributions, and, also to collect a few major facts about Fermi coherent states.

\begin{figure}
\begin{center}
\includegraphics[width=6cm]{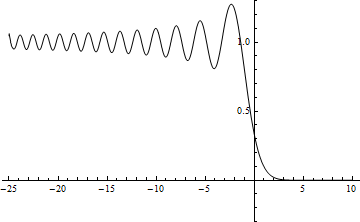}
\caption{Universal behavior of the Wigner function vs. momentum  for a concave Fermi surface obtained from (\ref{A}). \label{WignerAiry}}
\end{center}
\end{figure}

\paragraph*{2. Coherent Fermi states}
The formal  definition of a Fermi coherent states starts with  the  current algebra (see e.g., \cite{stone:book}).  To simplify the discussion and formulas we consider only one chiral (right)  part of the current algebra.

Current modes are Fourier harmonics of the electronic density $\rho(x)=\sum_{k>0} e^{\frac{i}{\hbar}k x} J_k$.
An electronic current mode
$ J_{k}=\sum_p c^\dag_{p}c_{p+ k}$ (we count electronic momentum from the Fermi momentum $p_F$),
creates a superposition of particle-hole excitations with  momentum  $ k$.  Positive modes annihilate the ground state, $|0\>$, a state where all momenta below $p_F$ are filled: $J_{k}|0\>=0,\quad k>0$.  Negative modes are Hermitian conjugated to the positive modes $J_{-k}=(J_{+k})^\dag$.

Chiral currents  obey a current  (or Tomonaga)  algebra:
\begin{align} \label{ca}
[J_k,\,J_{l}]= \frac k\Delta \delta_{k+l,0}.
\end{align}
A Fermi coherent state $|U\>$ is defined as an eigenstate of positive current modes:
\begin{align}\label{CoherentIsEigenstate}
J_{k}|U\rangle= p_{k}|U\>,\quad k>0.
\end{align}
As follows from (\ref{P}), $p_k= \Delta \int e^{-\frac{i}{\hbar} kx} d\Phi(x)/(2\pi)$ are positive Fourier modes of the function $\Phi (x)$. Assuming that  the total number of particles in the coherent state is the same as in the ground state, or that the dc component of current is $I_0=\frac{e v_F}{
\hbar}p_F$, i.e., $\int d\Phi=0$, we obtain
\begin{align}
|U\>=Z^{-1/2}e^{ \sum_{k>0} \frac{p_k}{k}J_{-k}}|0\>,\quad
Z=e^{\sum_{k>0}\frac 1k |p_k|^2}.
\end{align}
Using normal ordering with respect to the ground state (where all positive modes of the current are placed to the right  of negative modes) the unitary operator reads \footnote{The formula contains the physics of the Anderson orthogonality catastrophe \cite{Anderson:Catastrophe}: the  overlap between the ground state of a Fermi gas and a coherent state of a Fermi gas describing a  localized potential vanishes with a power of the level spacing. If the energy dependence of the  scattering phase $\delta$ of the potential is a smooth  at the Fermi energy  $p_k \to  \delta/\pi$ at small $k$.   As a result, the overlap $|\<0|U\>|^2=Z^{-1}=e^{\sum_{k=0}^{p_F}\frac {1}{ k}|p_k|^2}\sim (\Delta/p_F)^{(\delta/\pi )^2}$ vanishes with the spacing.}:
\begin{align}
&e^{i \int \Phi(x)\rho(x) dx}=Z^{-\frac12}\!:\!e^{ i \int \Phi(x)\rho(x) dx}\!:= \nonumber \\
&=Z^{-\frac12}e^{\sum_{k>0} \frac {1}{ k}p_kJ_{-k}}e^{- \sum_{k>0}  \frac {1}{ k}p_{-k}J_{k}}.
\end{align}

A coherent state represents an electronic   wave-packet which
is  fully   characterized by the electronic  density. It is a simple exercise involving the algebra of the current operators to show that the function $\nabla\Phi(x)$ is a non-uniform part of the density as is in (\ref{ro}). Alternatively, one can use $Z$ as a
generation function $\<\rho(x)\>\equiv \<U|\rho(x)|U\>=\rho_0+2{\rm Re}\sum_{k>0}ke^{ikx}\partial_{\bar p_k}\log Z$.

Coherent states obey the  Wick  theorem. The Wick theorem allows  to compute a correlation function of any finite number of electronic operators, as a determinant over the one-fermionic function $K(x_1,x_2)\equiv \<U|\psi^\dagger(x_1) \psi(x_2) |U\>$, where $\psi(x)=(\hbar/\Delta)^{1/2}\sum_p e^{\frac i\hbar px}c_p$ is an electronic operator. The one-fermionic function can be computed with the help of the formula:
\begin{align}\label{U}
U\psi(x)U^{-1}=e^{-i \Phi(x)}\psi(x)
\end{align}
which leads to the expression:
\begin{align}\label{G}
&K(x_1,x_2)=
\frac{e^{i\Phi(x_1)-i\Phi(x_2)-\frac i\hbar p_F(x_1-x_2)}-1}{i(x_1-x_2)},
\end{align}
valid for $\Delta |x_1-x_2| \ll \hbar$. An equivalent object appears in RMT where it is often called -  Dyson's kernel. We adopt this name.
As points merge one recovers the density (\ref{ro}) $K(x,x)=\<\rho(x)\>=\nabla\Phi$.

\paragraph*{3. Wigner function and Fermi occupation  number}

The Wigner function is defined as Wigner transform of the Dyson kernel \begin{align} \label{Wignerdefinition}
n_F(x,p) = \frac{1}{2\pi}
 \int K(x +\frac y2,x - \frac y2)e^{-\frac{i}{\hbar}py }dy
 \end{align}
The meaning of the Wigner function is clarified away from the surf. There  it means an occupation of electrons in the phase space $(x,p)$: 1 below a surf, 0 above. On the surf  Wigner function is not necessarily positive.

The Fermi number
\begin{align}\label{FD}
n_F(p)=
\< U|c^\dag_pc_p|U\>=\frac{\Delta}{2\pi \hbar}\int  n_F(x,p) dx
\end{align}
is the Wigner function averaged over space.

Combining    \eq{G} and \eq{Wignerdefinition} we write
\begin{align}\label{n}
n_F(x,p)=\frac{1}{2\pi i }\int e^{\frac{i}{\hbar}\int^{x+\frac y2}_{x-\frac y2} \left(P_F(x')-p\right)dx'} \frac{dy}{y-i0}.
\end{align}
where we denoted $P_F(x)=\hbar\nabla\Phi=\<\rho(x)\>$ as in  \eq{ro}.

Below we evaluate  the integral \eq{n}
semiclassically  bearing in mind that $\Phi$ is of a finite order as $\hbar\to0$.

A universal regime arises at the Fermi surf, $\kappa |p-P_F(x)| \simeq 1$. In this case it is sufficient to expand $P_F(x)$ in a Taylor series around extrema of $P_F(x)$  to second order $P_F(x)=P_F(x_*)+\frac 12 P_F''(x_*)(x-x_*)^2+\dots$.  Then the integral \eq{n} becomes the Airy integral given in Eq. (\ref{A}). Further integration over space yields (\ref{W}).

In this regime the Dyson kernel in   the momentum space $K_{p_1,p_2}\equiv\<U|c^\dag_{p_1}c_{p_2}|U\>$
reads:
\begin{align}\label{K}
K_{p_1,p_2}\approx \Delta \frac{{\rm Ai}( \kappa p_1){\rm Ai}'( \kappa p_2)
-{\rm Ai}( \kappa p_2){\rm Ai}'( \kappa p_1)}{ (p_1-p_2)}
\end{align}
This is the celebrated Airy kernel appearing in numerous problems as the limiting shape of crystals  \cite{Spohn}, asymmetric diffusion \cite{TASEP}, edge distribution of eigenvalues of random matrices \cite{Tracy:Widom}, etc.

The Fermi number (\ref{W}) can be directly  obtained from the kernel by taking a limit $p_1\to p_2$ in (\ref{K}). At large positive momenta ($\kappa p \to +\infty$) the Fermi number behaves as $n_F(p)\sim \frac{\Delta }{8\pi p}e^{-\frac 4 3(\kappa p)^{3/2}}$  and as $\sim \frac{\Delta}{\pi}\left( \kappa \sqrt{-p \kappa}-\frac{1}{4p}\cos(\frac 4 3(-\kappa p)^{3/2})\right)$  for large negative momenta within the surf.

Away from the universal region of the surf the Fermi distribution  can be computed within a saddle point approximation.  The saddle point of the integral \eq n is:
\begin{align}\label{sp}
P_F(x+\frac y 2)+P_F(x-\frac y 2)=2p
\end{align}
It  has pairs of solutions $\pm y_*(x,p)$.   Let  $P_{\rm max}=\max (P_F(x))$ and $P_{\rm min}=\min (P_F(x))$ be adjacent extrema of the surf. Without loss of generality we may assume that $(x,p)$ is outside the Fermi sea  $p>P_F(x)$.  The particle hole symmetry $n_F \to 1 - n_F$ helps to recover the case when the momentum is inside the sea.  If $p$ is in the surf, $p\in (P_{\rm min}, P_{\rm min})$, then   some saddle point pairs of \eq{sp}  may be real.  Their contribution produces oscillatory features with a suppressed  amplitude. If $p$ hovers above the surf, $p>P_{\rm max}$, then the saddle points are imaginary. Their  contributions are exponentially small.

Between two adjacent extrema  the Wigner function reads:
\begin{align} \nonumber
n_F(x,p)\approx
\sqrt{\frac{\hbar \left|\frac{dy_*}{dp} \right|}{8 \pi |y_*|^2}}\times
\left\{
\begin{array}{lr}
 2 \sin\left( \Omega-\frac\pi 4\right),&p\in(P_{\rm min}, P_{\rm max})\\
 e^{-|\Omega|},& p>P_{\rm max},\\
\end{array}
\right.
\end{align}
where
$\hbar \Omega=-\int_{P_F(x)}^py_*(x,p')dp'$.
In the surf it is half the action of a semiclassical periodic orbit - the area of the graph $y_*(x,p)$ vs $p$. In the universal regime, when one approximates
$y_*(x,p)\approx  \hbar \kappa^{3/2} \left (p-P_F(x)\right)^{1/2}$ this equation reproduces asymptotes of Eq.\eq{A}: $n_F(x,p+P_F(x))\sim (8\pi)^{-1/2}(\kappa p)^{-3/4}e^{-3(\kappa p)^{3/2}}$.

A Fermi coherent state with a periodic current is an instructive example.  It corresponds to "quantum pumping"  -  periodic transfer a charge through the system by applying a periodic voltage  through a point contact.  Setting $\<\rho (x)\>=\rho_0+(n/\ell)\cos(x/ \ell)$ the Dyson kernel in the momentum representation becomes the integer  Bessel kernel
\begin{align}\label{K1}
K_{p_1, p_2 }=\frac{n}{2}\frac{m_1 J_{m_1}(n)J_{m_2}'(n)-m_2 J_{m_2}(n)J_{m_1}'(n)}{m_1-m_2},
\end{align}
where $p_{1,2}=\hbar m_{1,2}/\ell$, where  $m_1 \neq m_2$ are integers. The Fermi number is given by ($m_F=p_F \ell/\hbar$):
\begin{align}\label{BesselFermiNumber}
n_F(p)= \frac{1}{2} - {\rm sign}(p) \left[\frac{J_0^2(n)}{2} +  \sum_{m=1}^{m_F}  J_m^2(n) \right].
\end{align}
This formula allows to  compare the asymptotes near the edges to the universal expression above. Using the homogeneous asymptote of Bessel function $J_m(m-(m/2)^{1/3}\zeta)\sim (2/m)^{1/3}{\rm Ai}(\zeta)$ at large $m$, one  recovers (\ref{K}) and (\ref{W}). Fig.~\ref{TheFigure} illustrates the universal asymptote.

\paragraph*{4. Holomorphic Fermions as  coherent states}
To contrast  semiclassical coherent Fermi states and quantum coherent Fermi states, we briefly discuss special coherent states known as holomorphic fermions  \cite{Miwa1}.

Holomorphic fermions are defined as a superposition of fermionic modes
 $\psi(z)=\sum_{p}e^{\frac i\hbar pz}c_p$,
 with a complex "coordinate" ${\rm Im} z < 0$.

Holomorphic fermions  are coherent states  since they can be represented as an exponent of a Bose field - displacement of electrons $\varphi(z)=\sum_{k\neq 0}\frac{\Delta}{i  k}e^{\frac{i}{\hbar}k z} J_k $ \cite{Miwa,stone:book}
 \begin{align}\label{f}\psi (z)=c_{p_F}
 :e^{i \varphi(z)}:\end{align}
A function $\Phi$ for a string of fermions  $\prod_{i=1}^n \left( \psi^\dag( z_i)\psi( \zeta_i)\right)|0\>$ is $e^{i\Phi(x)}=\prod_{i=1}^n\frac{x- z_i}{x-  \bar z_i}\frac{x-  \bar \zeta_i}{x-  \zeta_i}$. The density  (or current) of these states consists of Lorentzian  peaks, each carrying a unit electronic (positive/negative) charge $\<\rho(x)\>-\rho_0=\sum_{i}{\rm Im}\,\left(\frac{1}{x-z_i}- \frac{1}{x-  \zeta_i}\right)$, so that the state is  a set of single electronic pulses. For a possible applications of these states in nano-devices see \cite{Keeling,pointcontact}. As the complex coordinate approaches  the real axis $ \zeta=x-i0$ a holomorphic fermion operator  becomes an electronic operators $\psi(z)\to\psi(x)$ as its density becomes a delta-function.

Coherent states formed by a single holomorphic fermion carries a  unit charge  in contrast to semiclassical Fermi states. The Wigner function of this state
follows from  \eq{G}
\begin{align}\nonumber
n_F(x,p) =\frac{1}{2\pi i}\int \frac{x+\frac y2- z}{x+\frac y2- \bar z}\frac{x-\frac y2- \bar z}{x-\frac y2-z}e^{-\frac i\hbar py} \frac{dy}{y-i0},
\end{align}
where $z=X-ia$, $X$ is a real coordinate of fermion and $a$ is its width ($a\Delta \ll \hbar$) . $a$ is positive for an annihilation operator and negative for a creation operator.
Evaluating this integral we obtain
\begin{align}\nonumber
&n_F(x,p)=\Theta(-p)  - \Theta(-pa) a\frac{\sin 2p(x-X)}{x-X} e^{\frac{2}{\hbar}pa},\\
&n_F(p)=\Theta(-p) -  \Theta(-pa)\frac{2a\Delta}{\hbar} e^{\frac 2\hbar pa },
\end{align}
Noticeable features of this distribution are: the Fermi function jumps on the Fermi edge;
beyond the Fermi edge, the Fermi number and the Wigner functions decay exponentially, if $a>0$ a holomorphic fermion $\psi(z)$ acts like an annihilation operator removing particles from Fermi edge (vice versa if $a<0$); the Wigner function features Friedel's type oscillation with a distance $x-X$ to the center of the fermion.

\paragraph*{6. Fermi coherent states and Random Matrix Ensembles.}
We complete the letter by a brief discussion of the relation to the theory of Random Matrices.

Consider  $n$ electrons  from the ground state of the Fermi gas in positions $x_1,\dots,x_n$. We  write this state as $\<x_1,\dots,x_n|=\<0|c^\dag_{p_F}\dots c^\dag_{p_F-n}\psi (x_1)\dots\psi(x_n)$, where we set momentum spacing to 1 for brevity. Let us ask for the probability to find this state in the coherent state $|U\>$. It is  $|\Psi(x_1,\dots,x_n)|^2$, where $\Psi(x_1,\dots,x_n)=\<x_1,\dots,x_n|U\>/\<0|U\>$.

Fermions in coherent states obey the Wick theorem: a matrix element of a particle-hole string inserted between two (generally different) coherent states   is a Slater determinant  built out of particle-hole matrix element.
In particular  $ \Psi(x_1,\dots,x_n)={\rm det} \left( \Psi_{p_F-l}(x_m)\right)_{l,m\leq n}$ is a Slater determinant   of a single particle matrix element
$\Psi_{p_F}(x)=\<0|c^\dag_{p_F}\psi(x)|U\>/\<0|U\>$. We compute it with the  help of the current algebra and bosonic  representation \eq{f}. Up to a normalization
\begin{align}
\label{1p}\Psi_{p}(x)\sim e^{ -\frac 1\hbar Y_p(x)} .
\end{align}
This formula features the complex curve $Y_p(z)=-ip z+ \sum_{k>0}\frac{\hbar}{ k}p_ke^{\frac i\hbar kz}$, a useful characteristic of the coherent state. The function $Y_p(z)+ipz$ is analytic in the upper half-plane. Its   boundary value on the real axis is
\begin{align}
Y_p(x)+ipx=\frac{\hbar}{2}\left(V(x)-i\Phi(x)\right)=i\hbar\sum_{k>0}\frac{p_k}{k} e^{\frac{i}{\hbar} kx},
\end{align}
where $V(x)$ and $\Phi(x)$ are real and imaginary part of the boundary value of the analytic function. They  are connected by the Hilbert transform.

For example, a complex curve  for a pumping considered in  Sec. 3 is $Y_p(z)=-ipz+\frac {\hbar n}{2\ell}e^{\frac z\ell}$.
In the  case of a string of fermions the curve is $Y_p(z)=-ipz+ \hbar\sum_{i=1}^n\log \frac{z- \bar \zeta_i}{z-\bar z_i}$.

Computing the Slater determinant of \eq{1p} we obtain
 \begin{align}\label{first}
 \Psi(x_1,\dots,x_n)=Z_n^{-1/2}e^{ -\frac 1\hbar\sum_{i=1}^n(Y(x_i)-ip_Fx_i)}\Delta(x),
 \end{align}
  where
 $\Delta(x)=\mbox{det}\,\left(e^{\frac i\hbar p x_l}\right)_{p,l\leq n}=\prod_{i>j}\left(e^{\frac i\hbar x_i}-e^{\frac i\hbar x_j}\right)$ is  the VanderMonde determinant.

The normalization factor in (\ref{first})
 \begin{align}\label{first1}
Z_n=\int  \prod_{n\geq i>j\geq 1}|e^{\frac i\hbar x_i}-e^{\frac i\hbar x_j}|^2\prod_{i=1}^n e^{-  V(x_i)}dx_i
\end{align}
 is the the partition function of eigenvalues of a circular unitary $n\times n $-matrix \cite{Forrester-book}. At the limit of vanishing  spacing one replaces $e^{\frac i\hbar x}\to 1+i\hbar x$. In this case coherent Fermi state is described by Random Hermitian Matrix ensemble.

If $n$ is large, Eq. \eq{first} can be interpreted as a coordinate representation of the coherent state.
A Fermi coherent state  may be thought  as a Fermi sea filled  by  particles (without holes) with wave functions
 (\ref{1p})  and  $p=1,\dots,n$.  The coordinate representation provides another avenue to compute  matrix elements discussed in this paper as a limit $n\to\infty,\, p_F/n\to \infty$. Some of them have been studied for various reasons  in the theory of Random Matrix Ensembles (see e.g., \cite{Brezin:Zee} for derivation of the Dyson kernel).

\paragraph*{Acknowledgment}
 P. W. was supported by NSF DMR-0906427, MRSEC under DMR-0820054. E. B. was supported by grant 206/07 from the ISF.


\begin{thebibliography}{11}
\bibitem{Miwa1} In mathematical literature holomorphic fermions have been introduced in \cite{Miwa}.
For some physical applications including shot noise of these states, see recent papers \cite{{Keeling}}
\end{thebibliography}
\end{document}